\documentclass{vietnam}

\usepackage{slashed}
\usepackage{amsmath} 
\def\be{\begin{equation}}
\def\ee{\end{equation}}
\def\bea{\begin{eqnarray}}
\def\eea{\end{eqnarray}}

\newcommand{\Photo}{}

\begin{document}
\vspace*{4cm}
\title{Direct detection of boosted dark matter in two component dark matter scenario}

\author{ Keiko I. Nagao }

\address{Department of Physics, Okayama University of Science\\
1-1 Ridaicho Kita-ku, Okayama, Japan}

\maketitle
\abstracts{
We investigate the detection of boosted dark matter in a two component dark matter model with the hidden gauge $U(1)_D$ symmetry. 
The model introduces heavy and light fermionic dark matter components, where the heavy dark matter annihilation in the Galactic center boosts the light dark matter. 
Directional direct detection experiments, such as NEWSdm, are suitable for testing the boosted scenario by focusing on signals from the Galactic center. 
We found that the expected event rate could reach up to O(10) events per year per 5 kg of the detector material.}

\section{Introduction}
\footnote[0]{This work is based on collaboration with Tatsuhiro Naka and Takaaki Nomura \cite{thepaper}.}
Understanding the nature of dark matter (DM) is a big subject that spans astrophysics, cosmology and particle physics.
So far, many direct searches for Weakly Interacting Massive Particles (WIMPs) have been conducted, and the results have severely constrained its parameter space.
Light DM with a mass of sub-GeV is difficult to verify in direct detection experiments because its kinetic energy can be lower than the detector's energy thresholds. 
However, it can be tested if it is accelerated and has kinetic energies higher than the threshold.

In the boosted dark matter scenarios, dark matter with mass below the sub-GeV scale, which is lighter than the ordinary WIMP, is assumed to be boosted by some mechanisms like cosmic ray scattering, heavy particle annihilation, and decay into dark matter~\cite{Agashe:2014yua,Elor:2015tva,Aoki:2018gjf,Li:2023fzv}.
In this study, we consider a two component dark matter model in which the annihilation of the heavy dark matter boosts the light dark matter. 
Such boosting events would occur mainly in the center of the Milky Way galaxy, where dark matter density is high. 
Therefore, the boosted light dark matter scenario could be more suitably tested if we could examine the signal with directional direct detection that has a sensitivity to the direction of dark matter.

This paper is organized as follows: Section \ref{sec:Model} describes the two component dark matter model, Section 3 examines boosted dark matter events from the Galactic center with direction-sensitive dark matter detection, and Section 4 concludes.

\section{Model}
\label{sec:Model}

\begin{table}[h]
\caption[]{List of fields to be added to the standard model.}
\label{tab:fieldcontents}
\vspace{0.4cm}
\begin{center}
\begin{tabular}{c|ccc} \hline
   Fields           & $\psi$    & $\chi$    & $\varphi$ \\ \hline
   $U(1)_D$ charges & $Q_\psi$  & $Q_\chi$  & 1 \\ \hline
   Explanation & Heavy DM  & Light DM  &  \\ \hline
\end{tabular}
\end{center}
\end{table}

Let us introduce hidden $U(1)_D$ gauge symmetry and three additional fields to the Standard Model: fermion fields $\psi$ and $\chi$, and a scalar field $\varphi$.
Note that hidden $U(1)_D$ gauge symmetry also introduces a new gauge boson $A'$, which is a so-called dark photon. The dark photon obtains its mass after the scalar $\varphi$ acquires the vacuum expectation value.
Fields $\psi$, $\chi$ and $\varphi$ have $U(1)_D$ charges $Q_\psi$, $Q_\chi$ and $1$, respectively as listed in Tab. \ref{tab:fieldcontents}. 
Charges of the fermion fields should satisfy $|Q_\chi| \neq |Q_\psi |$, $|Q_\chi| \neq 1$, $| Q_\psi| \neq 1$ and $| Q_\chi \pm Q_\psi | \neq 1$ to forbid fermion mixing terms. 
Then we have remnant discrete symmetry $Z_2^\chi \times Z_2^\psi$ where both $\chi$ and $\psi$ are odd under $Z_2^\chi$ and $Z_2^\psi$, respectively. 
The discrete symmetry allows $\psi$ and $\phi$ to be dark matter.

Kinetic mixing of a dark photon with the standard model gauge field results in dark photon interaction with the standard model electromagnetic current $J_\mathrm{EM}^\mu$.
\begin{equation}
\mathcal{L}_{A'} = e \epsilon J^\mu_{\rm EM} A'_\mu,
\end{equation}
where $\epsilon$ is a tiny parameter associated with the kinetic mixing, and $e$ is the electromagnetic constant.

While, by introducing $\chi$ and $\phi$, the new Lagrangian terms are added to the Lagrangian
\begin{equation}
\mathcal{L}_{\rm DM} = \bar \chi (i \slashed{D} - m_\chi) \chi + \bar \psi (i \slashed{D} - m_\psi) \psi, 
\end{equation}
where $D_\mu \chi=(\partial_\mu + i Q_{\chi} g_D A'_\mu) \chi$ and $D_\mu \psi=(\partial_\mu + i Q_{\psi} g_D A'_\mu) \psi$ are the covariant derivatives, and $g_D$ is gauge coupling of hidden $U(1)_D$ gauge symmetry.
Thus, the interaction term of the dark photon with the new fermions are
\begin{equation}
A'_\mu (g_D Q_\chi \bar \chi \gamma^\mu \chi + g_D Q_\psi  \bar \psi \gamma^\mu\psi).
\end{equation}
Hereafter, they will be written as $g_\chi=g_D Q_\chi$ and $g_\psi=g_D Q_\psi$ for simplicity.

We suppose $\psi$ is heavier than $\chi$, i.e. $m_\chi < m_\psi$. 
The relic abundance of $\psi$ is determined from the freeze-out of annihilation process $\psi \psi \to A_\mu \to \chi \chi$. Its cross section is proportional to $g_\psi^2 g_\chi^2$. 
On the other hand, the relic abundance of $\chi$ is determined by the freeze-out of $\psi \psi \to A_\mu' A_\mu'$, its cross section proposes to $g_\psi^4$. 
If the couplings satisfy the condition $g_\psi < g_\chi$, the freeze-out of $\psi$ occurs earlier than the freeze-out of $\chi$. Therefore, the relic abundance of $\psi$ is larger than that of $\chi$, making $\psi$ the main component of dark matter. 
The value of coupling $g_\psi$ should be tuned so that the relic abundance of $\psi$ is consistent with the observations; $\Omega_\mathrm{DM}h^2\simeq 0.1$.

\section{Directional Detection of Dark Matter Boosted from the Galactic Center} \label{sec:detection}

The number of events caused by boosted $\chi$ from the direction of the Galactic center in the direct detection experiment can be estimated as 
\begin{equation}
N= \Delta T N_\mathrm{target}\Phi^{10^\circ}_\mathrm{GC} \int d E_R \frac{d \sigma_{\chi N\to \chi N}}{dE_R},
\label{eq:dNdER}
\end{equation}
where $\Delta T$ is exposure time, $N_\mathrm{target}$ is the number of targets in the detector, $\Phi^{10^\circ}_\mathrm{GC}$ is the $\chi$ flux from a 10$^\circ$ cone around the Galactic center, and $E_R$ is the recoil energy of target nucleus. The $\chi$ flux $\Phi^{10^\circ}_\mathrm{GC}$ can be roughly evaluated as 
\begin{equation}
\Phi^{10^\circ}_\mathrm{GC}=2.0\times10^{-2}\textrm{cm}^{-2}\textrm{s}^{-1}
C_{\rm pro}\left(\frac{\langle \sigma_{\psi\bar{\psi}\to \chi\bar{\chi}} v\rangle}{5\times 10^{-26} \ \textrm{cm}^3/\textrm{s}}\right)
\left(\frac{60\  \textrm{MeV}}{m_\psi}\right)^2, 
\label{eq:flux}
\end{equation}
where $C_{\rm pro}$ is a factor depending on DM profile. In this study, we assume the Einasto profile as DM profile; in that case, $C_{\rm pro}\simeq 3.76$.
The event number in the detector is represented as 
\begin{align}
\frac{dN}{d\ \log{E_R}}&=E_R\frac{dN}{dE_R}=\Delta T N_\mathrm{target}\Phi^{10^\circ}_\mathrm{GC} E_R \frac{d \sigma_{\chi N\to \chi N}}{dE_R} \nonumber,\\
N&= \Delta T N_\mathrm{target}\Phi^{10^\circ}_\mathrm{GC} \int d E_R \frac{d \sigma_{\chi N\to \chi N}}{dE_R}.
\label{eq:dNdER}
\end{align}

\begin{figure}[tb]
 \begin{center}
\includegraphics[width=7cm]{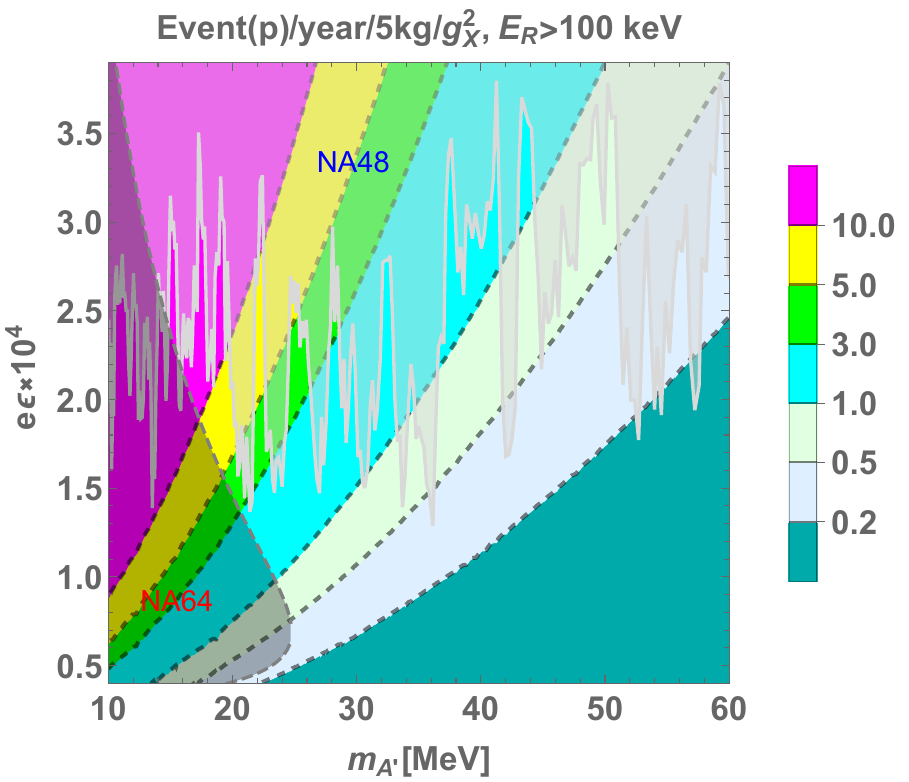} \ 
\includegraphics[width=7cm]{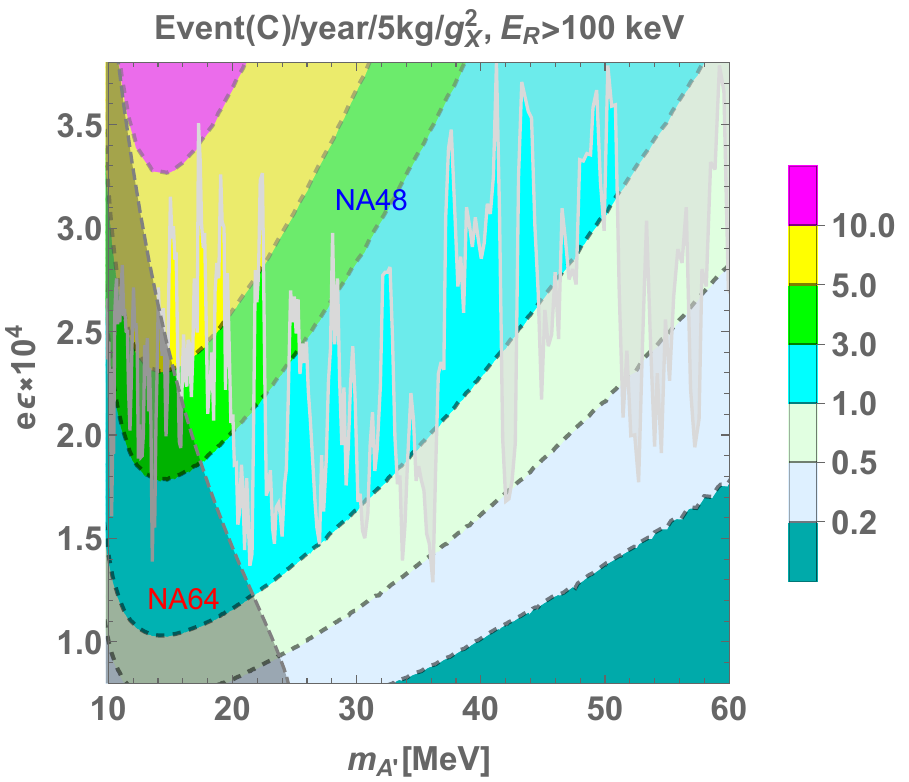} \\ 
\includegraphics[width=7cm]{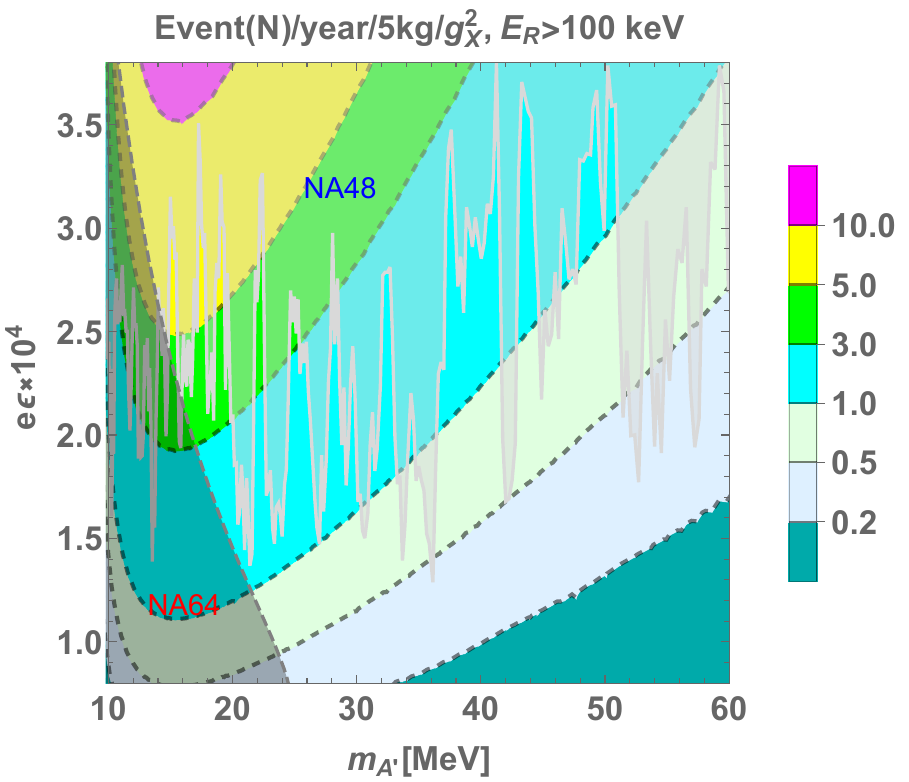} \ 
\includegraphics[width=7cm]{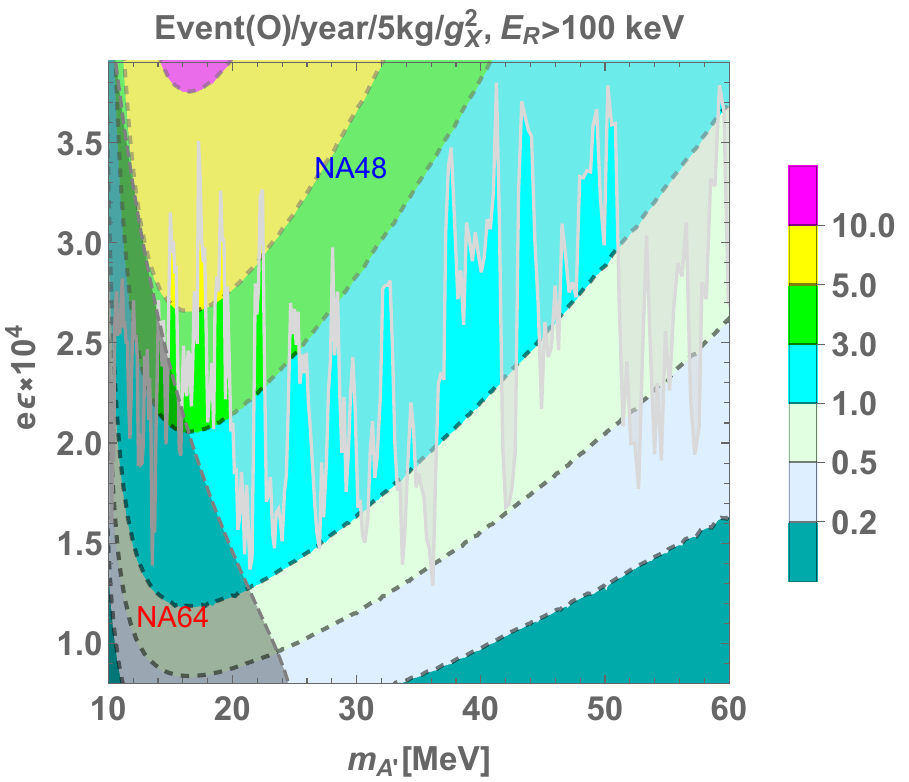} 
\caption{Expected event number in NEWSdm experiment, i.e., the targets are $p$, $C$, $N$, and $O$. Detector thresholds are conservatively estimated to be currently used in detectors.}
\label{fig:Events2}
\end{center}
\end{figure}

\begin{figure}[tb]
 \begin{center}
\includegraphics[width=7cm]{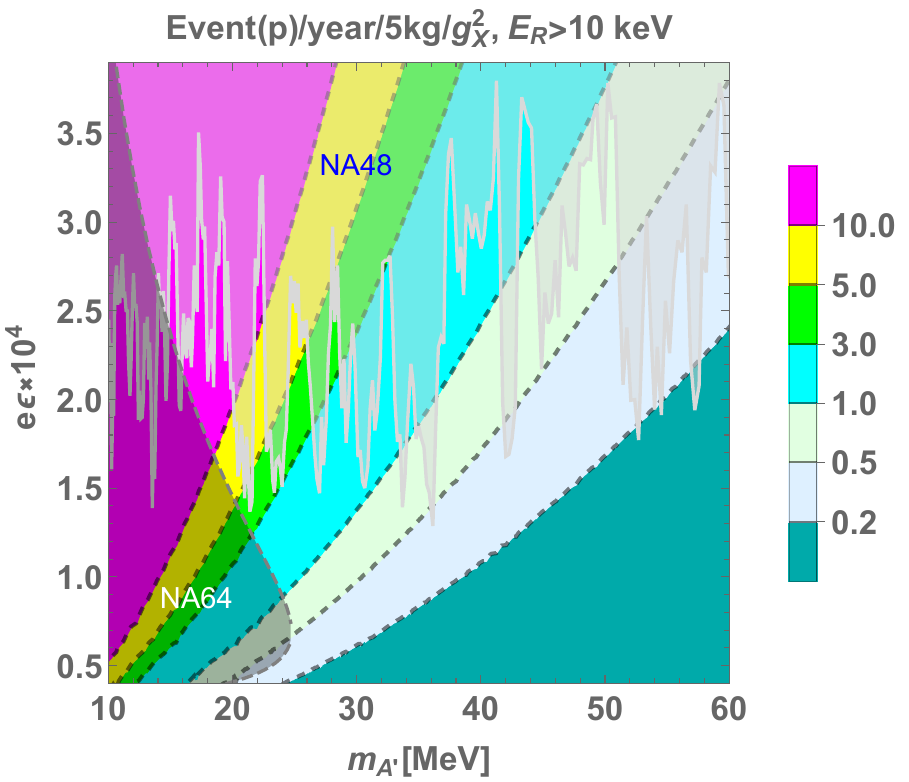} \ 
\includegraphics[width=7cm]{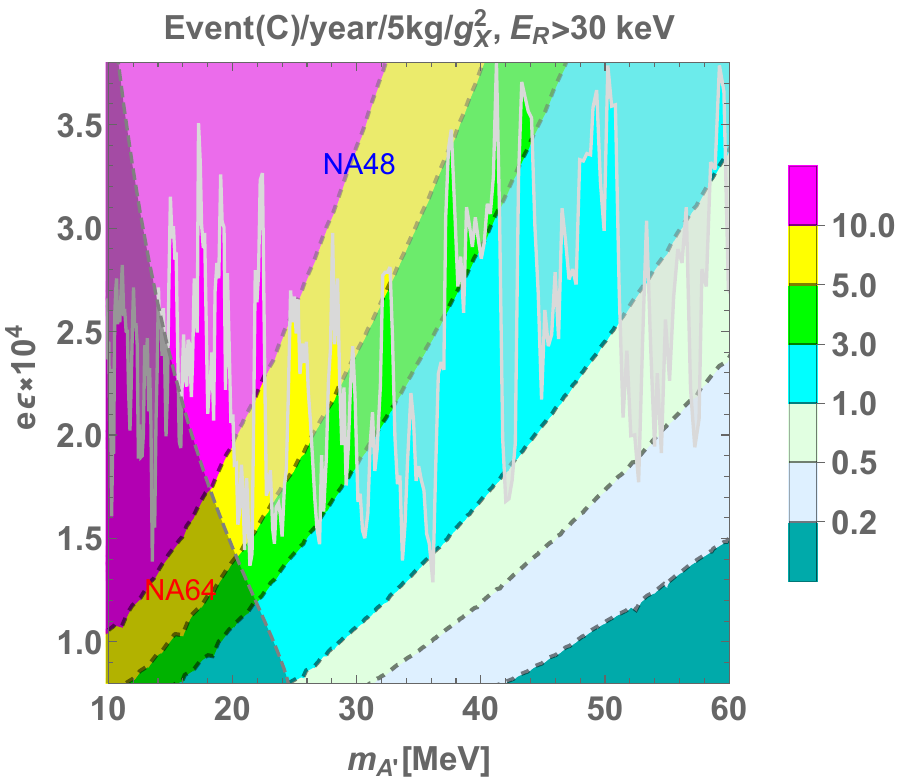} \\ 
\includegraphics[width=7cm]{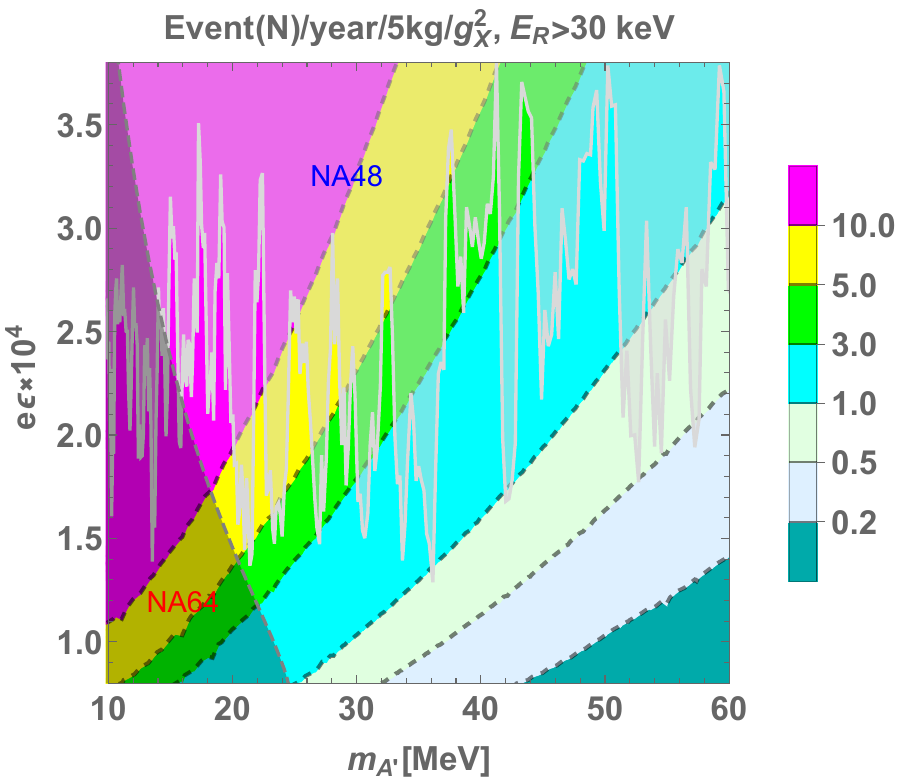} \ 
\includegraphics[width=7cm]{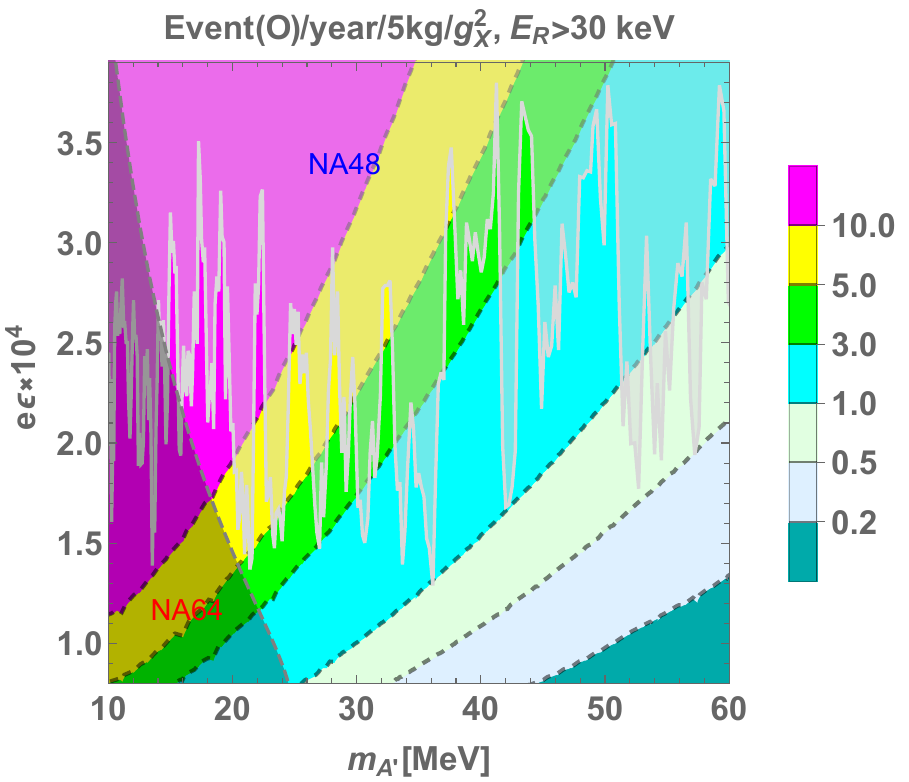} 
\caption{The legend is same as Fig.\ref{fig:Events2} except for the energy threshold of the detector. Detector thresholds are assumed based on what is expected to be achievable.}
\label{fig:Events1}
\end{center}
\end{figure}

The estimations of the event number are shown in Figure~\ref{fig:Events2} for the realistic energy threshold of the detector and Figure~\ref{fig:Events1} for the optimistic one. The targets in detectors are assumed to be relatively light atoms in super fine grained nuclear emulsion used in NEWSdm experiment, $p$, $C$, $N$, and $O$~\cite{NEWSdm:2024}. 
For simplicity, the mass relations $m_\psi=3m_\chi, m_\chi=m_{A'}$ are supposed.

\section{Conclusion}
We study the model with hidden $U(1)_D$ gauge symmetry, which has two dark matter components.
Heavy dark matter annihilation in the Galactic center will boost the light dark matter component.
Thus, we estimate the event number detected in a directional detection  NEWSdm, which is one of suitable detectors for testing the model. 
At most, the estimated event numbers are $\sim$ 10/year/5kg/$g_\chi^2$. 

In the specific case where the masses of the two dark matter components are degenerate, detection becomes more challenging because the boost to the lighter dark matter is smaller.
However, since the lightest target $p$ can be scattered even if the momentum of boosted DM is as tiny as $\sim O(1)$ MeV, detection of boosted light dark matter may still be possible. 
For details, see the main paper \cite{thepaper}.

\section*{Acknowledgments}
The work is supported by JSPS Grant-in-Aid for Scientific Research (C) 21K03562, (C) 21K03583, (C) 24K07061, (C) 24K07032 and Grant-in-Aid for Transformative Research Areas (A) 24H02244.



\section*{References}

\begin{thebibliography}{99}
\bibitem{thepaper} 
K.~I.~Nagao, T.~Naka and T.~Nomura,
[arXiv:2411.10149 [hep-ph]].

\bibitem{Agashe:2014yua}
K.~Agashe, Y.~Cui, L.~Necib and J.~Thaler,
JCAP \textbf{10} (2014), 062
doi:10.1088/1475-7516/2014/10/062
[arXiv:1405.7370 [hep-ph]].

\bibitem{Elor:2015tva}
G.~Elor, N.~L.~Rodd and T.~R.~Slatyer,
Phys. Rev. D \textbf{91} (2015), 103531
doi:10.1103/PhysRevD.91.103531
[arXiv:1503.01773 [hep-ph]].

\bibitem{Aoki:2018gjf}
M.~Aoki and T.~Toma,
JCAP \textbf{10} (2018), 020
doi:10.1088/1475-7516/2018/10/020
[arXiv:1806.09154 [hep-ph]].

\bibitem{Li:2023fzv}
J.~Li, T.~Nomura, J.~Pei, X.~Yin and C.~Zhang,
Phys. Rev. D \textbf{108} (2023) no.3, 035021
doi:10.1103/PhysRevD.108.035021
[arXiv:2302.09839 [hep-ph]].

\bibitem{NEWSdm:2024}
Tatsuhiro Naka and Giovanni de Lellis 
Journal of Advanced Instrumentation in Science, \textbf{2024}, (2024) 1
DOI: https://doi.org/10.31526/jais.2024.500
\end{thebibliography}
\end{document}